\documentclass[%
print,
 amsmath,amssymb,
 aps,
]{revtex4}

\usepackage[sc]{mathpazo}%此指令是令文章字体选用Palation.
\usepackage{graphicx}% Include figure files
\usepackage{dcolumn}% Align table columns on decimal point
\usepackage{bm}% bold math
\usepackage{dsfont}
\usepackage[landscape,papersize={297.1mm,210mm},left=1.9cm,right=1.6cm,top=2.7cm,bottom=2.8cm]{geometry}% 此指令是规定A4纸的页边距
\usepackage[                   %dvipdfm, pdflatex,pdftex这里决定运行文件的方式不同
            pdfstartview=FitH,
            colorlinks, %注释掉此项则交叉引用为彩色边框(将colorlinks和pdfborder同时注释掉)
            pdfborder=001,   %注释掉此项则交叉引用为彩色边框
            linkcolor=blue,
            anchorcolor=blue,
            citecolor=blue,
            urlcolor=blue
            ]{hyperref}
\usepackage{amsthm}
\makeatletter

\newcommand{\Rmnum}[1]{\expandafter\@slowromancap\romannumeral #1@}
\makeatother

\begin{document}
\title{Entanglement resource theory of quantum channel}

\author{Huaqi Zhou$^1$}

\author{Ting Gao$^1$}
\email{gaoting@hebtu.edu.cn}

\author{Fengli Yan$^2$}
\email{flyan@hebtu.edu.cn}
\affiliation{$^1$ School of Mathematical Sciences,
 Hebei Normal University, Shijiazhuang 050024, China \\
$^2$ College of Physics, Hebei Key Laboratory of Photophysics Research and Application, Hebei Normal University, Shijiazhuang 050024, China}

\begin{abstract}
Quantum channels  can represent dynamic resources, which are indispensable elements in many physical scenarios. To describe certain facets of nonclassicality of the channels, it is necessary to quantify their properties. In the framework of resource theory of quantum channel, we show two general ways of constructing entanglement measure of channels. We also present several entanglement measures of channels based on the Choi relative entropy of channels, concurrence and $k$-ME concurrence and give some specific examples. These entanglement measures of channels can deepen the cognizing about channel and advance the research on the transformation between  coherent resources and entangled resources. In addition, we prove that these measures satisfy the properties including nonnegativity, monotonicity, convexity and so on.\\

\end{abstract}


\maketitle

\section{Introduction}
Entanglement is an essential constituent part in quantum information and at the same time a striking feature of quantum mechanics \cite{Horodecki}. As a kind of special resource, it plays a significant role in quantum state discrimination \cite{40,43,20}, quantum teleportation \cite{PRL70.1895,PR448.1,18}, quantum key distribution \cite{Science283.2050, PRL67.661,PRL68.557,PR448.1}, etc. It is interesting work to quantify entanglement not only in theoretical research but also in practical application. Entanglement theory has already been carefully studied. Several entanglement measures have been introduced, such as relative entropy of entanglement \cite{57,78}, concurrence \cite{80}, $k$-ME concurrence \cite{86}, negativity \cite{58,65}, entanglement of formation \cite{54,246}. Most of the entanglement measures satisfy the entanglement monotone, which means that it does increase on average under local operations and classical communication. A few  better measures can satisfy the other important characteristics of an entanglement measure, including the convexity, subadditivity,  strong monotonicity.

The resource theory aims at providing a strict framework to qualify and quantify resource and, ultimately, to understand fully its capabilities and limitations within the field of quantum technologies \cite{Baumgratz, Chitambar, 122}. A powerful framework known as the entanglement resource theory mainly consists of three ingredients: separable states, free operations and entanglement measures \cite{41}. Separable states are those quantum states that do not possess any entanglement resource. It means that, in a multipartite system, measuring one or more of the parties does not have any effect on the particles in the other subsystems. Free operations are those quantum operations which do not generate entanglement from separable states. Entanglement measures being used to quantify the entanglement of quantum states are  the appropriate functions which map quantum states to real numbers. Concurrence and negativity mainly measure entanglement in the bipartite systems. Relative entropy of entanglement, $k$-ME concurrence and entanglement of formation can quantify entanglement in the multipartite systems.

The realization of quantum technology relates not only to quantum states but also to quantum channels, i.e. dynamical resources \cite{37}. On the level of entanglement, some channels can be naturally regarded as a kind of specially generalized resource, because of their ability to generate and increase the entanglement of quantum states. So we can establish entanglement resource theory of channels. It also needs three elements: free channels, free superchannels and entanglement measures of quantum channels. The channel which  converts separable states into separable states is called free channel. Here the superchannel is a map between channels \cite{38,123,52}. Free superchannels are those superchannels which transform free channels to free channels. The entanglement measures of channels are actually reasonable functions that sort the channels according to their entanglement and give the channels corresponding real numbers.

The quantification of resource is fundamental and important. It gives people a deeper understanding of basic physics and provides new insights and mathematical tools for various quantum information processing tasks. We know that certain quantum channels can be used to efficiently convert coherent states into entangled states. Therefore, the entanglement resource theory  for quantum channels is of great practical interest \cite{38}. For the measure of quantum channels, there are two methods can be taken, one is to think directly about the distance between the quantum channel and all the free channels, the other is to study free state conversion under certain quantum channel. The establishment of channel resource theory enhances and advances the research in the core field of quantum information. Some works on resource theories of quantum channels have been done \cite{41,100,35,37,38,52,181,125,123}. Among them, Ref. \cite{52}, Gour et al. defined the relative entropy of dynamical entanglement for bipartite entanglement and made a more detailed discussion on entanglement of bipartite POVMs. In Ref. \cite{181}, B$\ddot{\textup{a}}$uml et al. provided several entanglement measures for bipartite channels including the logarithmic negativity and the $\kappa$-entanglement.

In this work, we provide the Choi relative entropy of channels and some properties that it satisfies. Furthermore, we propose several entanglement measures of channels based on Choi relative entropy of channels, concurrence and $k$-ME concurrence respectively. The measure $\Re_{r}$ based on relative entropy can be considered in the multipartite system. We demonstrate that it is additivity under certain circumstances in addition to satisfying some basic properties containing nonnegativity, monotonicity and convexity. The measure $\Re_{c}$ based on concurrence being mainly discussed in bipartite systems has been investigated. We prove that this measure satisfies these fundamental properties. At last we study the measure $\Re_{k\textup{-ME}}$ based on $k$-ME concurrence. The $k$-ME concurrence is a generalized concurrence as a entanglement measure for multipartite finite-dimensional system. We also define the strength values of the channels. The purpose is to roughly quantify the ability of channels to convert fully separable states into $k$-nonseparable. The measure $\Re_{k\textup{-ME}}$ gives us a better understanding of the capability of channels. Moreover, we give several specific examples to compute the entanglement of channels.

\section{Entanglement resource theory of quantum channels}
Let us first introduce the concept of fully separable states. Consider a multiparticle  Hilbert space $\mathcal{H}=\mathcal{H}_1\otimes{\mathcal{H}_2}\otimes\cdots\otimes{\mathcal{H}_n}$, a quantum pure state $|\psi\rangle$ is called fully separable state if the pure state can be written as $|\psi\rangle=|\psi\rangle^1\otimes{|\psi\rangle^2}\otimes\cdots\otimes{|\psi\rangle^n}$, $|\psi\rangle^i\in{\mathcal{H}_i}$. A quantum mixed state $\rho$ is called fully separable state if the mixed state can be written as the convex combination of fully separable pure states, i.e., $\rho=\sum_{i}p_{i}\rho_{i}^{1}\otimes{\rho_{i}^{2}}\otimes\cdots\otimes{\rho_{i}^{n}}$ and $\rho_{i}^{j}\in{\mathcal{H}_{j}}$, $p_{i}\geq 0$, $\sum_{i}p_{i}=1$. Otherwise the state is called entangled state. Let $F$ express the set of all fully separable states in $\mathcal{H}$, $L(\mathcal{H})$ is the set of all the quantum channels from $\mathcal{H}$ to $\mathcal{H}$.

Furthermore we introduce the following symbols. $D(\mathcal{H})$ is the set of all quantum states in Hilbert space $\mathcal{H}$, $P(\mathcal{H})$ represents the set of all quantum pure states in Hilbert space $\mathcal{H}$. $\Gamma$ is indicated as the set of free channels $(\textup{i.e.}~\Gamma=\{N|N:F\rightarrow F\})$. $\widetilde{\Theta}$ denotes the set of superchannels. Here, superchannel $\Phi$ is a map which transforms a channel $N$ to another channel $\Phi(N)$ by $\Phi(N)=WNV$, where $W,V$ are also quantum channels in $L(\mathcal{H})$. We use $\Theta$ to express the set of free superchannels $(\textup{i.e.}~\Theta=\{\Phi_{\Gamma}\in \widetilde{\Theta}|\Phi_{\Gamma}:\Gamma\rightarrow\Gamma\})$.

\subsection{The measure based on the Choi relative entropy}
In probability or information theory, relative entropy is referred to as Kullback-Leibler divergence. It is used to measure the difference between two probability distributions. Similarly, the relative entropy of quantum states is to measure the difference between two states. Further, to measure the difference between two channels, the relative entropy of quantum channels has been proposed in Ref. \cite{97,17}. In a similar way, the Choi relative entropy of quantum channels is defined as
\begin{equation}
\begin{aligned}
S_{C}(N||M)=\max\limits_{X\in L(\mathcal{H})}S(N\cdot X\otimes \mathbb{I}(|\psi\rangle^{+}\langle \psi|)||M\cdot X\otimes \mathbb{I}(|\psi\rangle^{+}\langle \psi|)),
\end{aligned}
\end{equation}
where $N,~M$ are arbitrary two quantum channels belonging to $L(\mathcal{H})$, $|\psi\rangle^{+}$ is the maximally entangled state. $S(\rho'||\sigma')=\textup{Tr}[\rho'\textup{log}_{2}(\rho'/\sigma')]$ is quantum relative entropy \cite{87} for $\rho',\sigma'\in D(\mathcal{H})$. The maximum is taken over all quantum channels in $L(\mathcal{H})$. Note that the Choi relative entropy of channels, does not actually satisfy the usual metric properties, although known as a distance, e.g. $S_{C}(N||M)\neq S_{C}(M||N)$ in general case. In fact, it also can be written as
\begin{equation}
\begin{aligned}
S_{C}(N||M)=\max\limits_{X\in L(\mathcal{H})}S_{\Phi^{+}}(NX||MX).
\end{aligned}
\end{equation}
The relative entropy of Choi states of channels $S_{\Phi^{+}}=S(N\otimes \mathbb{I}(|\psi\rangle^{+}\langle \psi|)||M\otimes \mathbb{I}(|\psi\rangle^{+}\langle \psi|))$ comes from the Ref. \cite{17}. Generally, it satisfies the following properties.

$(i)$ $(Nonnegativity)$ $S_{C}(N||M)\geq0$ for any $N,M\in{L(\mathcal{H})}$.

$(ii)$ $(Monotonicity)$ $S_{C}(\Phi(N)||\Phi(M))\leq S_{C}(N||M)$ for any superchannel $\Phi$.

$Proof.$ By the definition, we get
\begin{equation*}
\begin{aligned}
S_{C}(\Phi(N)||\Phi(M))&=\max\limits_{X\in L(\mathcal{H})}S(\Phi(N)\cdot X\otimes \mathbb{I}(|\psi\rangle^{+}\langle \psi|)||\Phi(M)\cdot X\otimes \mathbb{I}(|\psi\rangle^{+}\langle \psi|))\\
                       &=\max\limits_{X\in L(\mathcal{H})}S(WNV\cdot X\otimes \mathbb{I}(|\psi\rangle^{+}\langle \psi|)||WMV\cdot X\otimes \mathbb{I}(|\psi\rangle^{+}\langle \psi|))\\
                       &\leq\max\limits_{X\in L(\mathcal{H})}S(WN\cdot X\otimes \mathbb{I}(|\psi\rangle^{+}\langle \psi|)||WM\cdot X\otimes \mathbb{I}(|\psi\rangle^{+}\langle \psi|))\\
                       &\leq\max\limits_{X\in L(\mathcal{H})}S(N\cdot X\otimes \mathbb{I}(|\psi\rangle^{+}\langle \psi|)||M\cdot X\otimes \mathbb{I}(|\psi\rangle^{+}\langle \psi|))\\
                       &=S_{C}(N||M).
\end{aligned}
\end{equation*}
Here the first inequality follows from that the set $L(\mathcal{H})$ contains the set $V\cdot L(\mathcal{H})=\{VX|X\in L(\mathcal{H})\}$, the second inequality is due to the monotonicity of relative entropy.

$(iii)$ $(Joint$ $convexity)$ $S_{C}(\sum\limits_{i}p_{i}N_{i}||\sum\limits_{i}p_{i}M_{i})\leq \sum\limits_{i}p_{i}S_{C}(N_{i}||M_{i})$ for any $N_{i},M_{i}\in{L(\mathcal{H})}$ with $p_{i}\geq 0$ and $\sum_{i}p_{i}=1$.

$Proof.$ Suppose the maximization is achieved with $X'$
\begin{equation*}
\begin{aligned}
S_{C}(\sum\limits_{i}p_{i}N_{i}||\sum\limits_{i}p_{i}M_{i})&=S(\sum\limits_{i}p_{i}N_{i}\cdot X'\otimes \mathbb{I}(|\psi\rangle^{+}\langle \psi|)||\sum\limits_{i}p_{i}M_{i}\cdot X'\otimes \mathbb{I}(|\psi\rangle^{+}\langle \psi|))\\
&\leq\sum\limits_{i}p_{i}S(N_{i}\cdot X'\otimes \mathbb{I}(|\psi\rangle^{+}\langle \psi|)||M_{i}\cdot X'\otimes \mathbb{I}(|\psi\rangle^{+}\langle \psi|))\\
&\leq\sum\limits_{i}p_{i}\max\limits_{X\in L(\mathcal{H})}S(N_{i}\cdot X\otimes \mathbb{I}(|\psi\rangle^{+}\langle \psi|)||M_{i}\cdot X\otimes \mathbb{I}(|\psi\rangle^{+}\langle \psi|))\\
&=\sum\limits_{i}p_{i}S_{C}(N_{i}||M_{i}),
\end{aligned}
\end{equation*}
where the first inequality results from the joint convexity of relative entropy, the second inequality is obvious.

$(iv)$ $(Additivity)$ $S_{C}(N_{0}\otimes N_{1}||M_{0}\otimes M_{1})\geq S_{C}(N_{0}||M_{0})+S_{C}(N_{1}||M_{1})$.

$Proof.$ Apparently, we have
\begin{equation*}
\begin{aligned}
S_{C}(N_{0}\otimes N_{1}||M_{0}\otimes M_{1})&=\max\limits_{X\in L(\mathcal{H}^{\otimes 2})}S((N_{0}\otimes N_{1})\cdot X\otimes \mathbb{I}(|\psi\rangle^{+}\langle \psi|)||(M_{0}\otimes M_{1})\cdot X\otimes \mathbb{I}(|\psi\rangle^{+}\langle \psi|))\\
&\geq\max\limits_{X_{0}\otimes X_{1}\in L(\mathcal{H}^{\otimes 2})}S((N_{0}\otimes N_{1})\cdot (X_{0}\otimes X_{1})\otimes \mathbb{I}(|\psi\rangle^{+}\langle \psi|)||(M_{0}\otimes M_{1})\cdot (X_{0}\otimes X_{1})\otimes \mathbb{I}(|\psi\rangle^{+}\langle \psi|))\\
&=\max\limits_{X_{0}\otimes X_{1}\in L(\mathcal{H}^{\otimes 2})}S((N_{0}X_{0}\otimes N_{1}X_{1})\otimes \mathbb{I}(|\psi\rangle^{+}\langle \psi|)||(M_{0}X_{0}\otimes M_{1}X_{1})\otimes \mathbb{I}(|\psi\rangle^{+}\langle \psi|))\\
&=\max\limits_{X_{0}\otimes X_{1}\in L(\mathcal{H}^{\otimes 2})}S_{\Phi^{+}}(N_{0}X_{0}\otimes N_{1}X_{1}||M_{0}X_{0}\otimes M_{1}X_{1})\\
&=\max\limits_{X_{0}\otimes X_{1}\in L(\mathcal{H}^{\otimes 2})}(S_{\Phi^{+}}(N_{0}X_{0}||M_{0}X_{0})+S_{\Phi^{+}}(N_{1}X_{1}||M_{1}X_{1}))\\
&=S_{C}(N_{0}||M_{0})+S_{C}(N_{1}||M_{1}),
\end{aligned}
\end{equation*}
where the inequality is obvious, the fourth equation arises from the additivity of the relative entropy $S_{\Phi^{+}}$ in Ref. \cite{17}.

In entanglement resource theory of channels, more concretely, we will build reasonable function of channels by taking into account the difference between the channels and all the free channels to directly quantify their value. Actually, the entanglement measure can be viewed as to a measure of entanglement resource generating and increasing capability of channels.

Definition $1$. For any quantum channel $N\in L(\mathcal{H})$, the entanglement measure of channels based on the Choi relative entropy of channels is defined as
\begin{equation}\label{3}
\begin{aligned}
\Re_{r}(N)=\min\limits_{M\in\Gamma}S_{C}(N||M).
\end{aligned}
\end{equation}

Note that this is a direct generalization of the relative entropy of entanglement. It is necessary to consider whether the measure $\Re_{r}$ has the similar properties. According to the features of relative entropy of channels, we have the following conclusions.

$(S1)$ $(Nonnegativity)$ $\Re_{r}(N)\geq 0$, for any $N\in L(\mathcal{H})$; and $\Re_{r}(N)=0$ if and only if $N\in\Gamma$.

$(S2)$ $(Weak$ $monotonicity)$ $\Re_{r}(\Phi_{\Gamma}(N))\leq\Re_{r}(N)$, for any $\Phi_{\Gamma}\in\Theta$.

$Proof.$ By the definition, we have
\begin{align*}
\Re_{r}(\Phi_{\Gamma}(N))&=\min\limits_{M\in\Gamma}S_{C}(\Phi_{\Gamma}(N)||M)\leq\min\limits_{M\in\Gamma}
S_{C}(\Phi_{\Gamma}(N)||\Phi_{\Gamma}(M))\leq\min\limits_{M\in\Gamma}S_{C}(N||M)=\Re_{r}(N),
\end{align*}
where the first inequality follows from that the set $\Gamma$ contains the set $\Phi_{\Gamma}(\Gamma)=\{\Phi_{\Gamma}(M)|M\in\Gamma\}$, the second inequality comes from the monotonicity $(ii)$ of Choi relative entropy of channels.

$(S3)$ $(Convexity)$ $\Re_{r}(\sum\limits_{i}p_{i}N_{i})\leq\sum\limits_{i}p_{i}\Re_{r}(N_{i})$, where $N_{i}$ is a quantum channel belonging to $L(\mathcal{H})$, and $\{p_{i}\}$ is a probability distribution satisfying $p_{i}\geq 0$ and $\sum_{i}p_{i}=1$.

$Proof.$ We find
\begin{align*}
\Re_{r}(\sum\limits_{i}p_{i}N_{i})&=\min\limits_{M\in\Gamma}S_{C}(\sum\limits_{i}p_{i}N_{i}||M)\\
                                  &\leq\min\limits_{\{M_{i}\}\subset\Gamma}S_{C}(\sum\limits_{i}p_{i}N_{i}||\sum\limits_{i}p_{i}M_{i})\\
                                  &\leq\min\limits_{\{M_{i}\}\subset\Gamma}\sum\limits_{i}p_{i}S_{C}(N_{i}||M_{i})\\
                                  &=\sum\limits_{i}p_{i}\min\limits_{\{M_{i}\}\subset\Gamma}S_{C}(N_{i}||M_{i})\\
                                  &=\sum\limits_{i}p_{i}\Re_{r}(N_{i}).
\end{align*}
Here we have utilized that a convex combination of free channels still is a free channel, and the set $\Gamma$ contains the set $\Sigma=\{\sum\limits_{i}p_{i}M_{i}|\{M_{i}\}\subset\Gamma\}$ in the first inequality, while the second inequality is due to the joint convexity $(iii)$ of Choi relative entropy of channels. The second equation is obtained from the nonnegativity $(i)$ of Choi relative entropy of channels and the mutual independence of choosing $M_{i}$.

$(S4)$ $(Strong$ $monotonicity)$ $\sum\limits_{i}p_{i}\Re_{r}(N_{i})\leq\Re_{r}(N)$, where $N_{i}=\Phi_{\Gamma}^{i}(N)$, $\Phi_{\Gamma}^{i}\in\Theta$, and $\{p_{i}\}$ is a probability distribution satisfying $p_{i}\geq 0$ and $\sum_{i}p_{i}=1$.

$Proof.$ According to the definition, we have
\begin{align*}
\sum\limits_{i}p_{i}\Re_{r}(N_{i})&=\sum\limits_{i}p_{i}\min\limits_{M\in\Gamma}S_{C}(N_{i}||M)\\
                                  &=\sum\limits_{i}p_{i}\min\limits_{M\in\Gamma}S_{C}(\Phi_{\Gamma}^{i}(N)||M)\\
                                  &\leq\sum\limits_{i}p_{i}\min\limits_{M\in\Gamma}S_{C}(N||M)\\
                                  &=\Re_{r}(N).
\end{align*}
Here the inequality follows from the weak monotonicity $(S2)$ of entanglement measure $\Re_{r}$.

$(S5)$ $(Additivity~under~restricted~condition)$ $\Re_{r}'(N_{0}\otimes N_{1})\geq \Re_{r}(N_{0})+\Re_{r}(N_{1})$, for any $N_{0},N_{1}\in L(\mathcal{H})$, where $\Re_{r}$ is denoted as $\Re_{r}'$ when $\Gamma$ is replaced by $\Gamma'$, and $\Gamma'$ is the set of all local free channels in space $\mathcal{H}\otimes\mathcal{H}$.

$Proof.$ Evidently, one has
\begin{align*}
\Re_{r}'(N_{0}\otimes N_{1})&=\min\limits_{M\in\Gamma'}S_{C}(N_{0}\otimes N_{1}||M)\\
                           &=\min\limits_{M_{0}\otimes M_{1}\in\Gamma'}S_{C}(N_{0}\otimes N_{1}||M_{0}\otimes M_{1})\\
                           &\geq \min\limits_{M_{0}\otimes M_{1}\in\Gamma'}[S_{C}(N_{0}||M_{0})+S_{C}(N_{1}||M_{1})]\\
                           &= \min\limits_{M_{0}\in\Gamma}S_{C}(N_{0}||M_{0})+\min\limits_{M_{1}\in\Gamma}S_{C}(N_{1}||M_{1})\\
                           &=\Re_{r}(N_{0})+\Re_{r}(N_{1}),
\end{align*}
where the inequality is due to the additivity $(iv)$ of Choi relative entropy of channels, and the third equality comes from the nonnegativity $(i)$ of Choi relative entropy of channels and the mutual independence of choosing $M_{i}$.

Because of the nature of the channel itself, an entanglement measure of channels also has some conditions and limitations. According to the definition of free channel, any non-free channel can always increase the entanglement of at least one quantum state. It is not difficult to see that the measure $\Re_{r}$ conforms to this case perfectly from conclusion $(S1)$. The conclusion $(S2)$ indicates that the entanglement of channels cannot increase under any free superchannel. These illustrate that the measure $\Re_{r}$ meets the basic requirements. Moreover, the conclusions $(S3)$-$(S5)$ show that it has some better properties.

\subsection{The measure based on the concurrence}
The concurrence was introduced by Wootters \cite{80,42} to measure entanglement of 2-qubit states. The concurrence for a 2-qubit pure state is defined by using the qubit spin-flip operator, and it is generalized to mixed states as the convex roof. Whereafter, the concurrence was extended to higher dimension cases \cite{21,22}. In the bipartite system $\mathcal{H}=\mathcal{H}_{1}\otimes \mathcal{H}_{2}$, the concurrence of a pure state $|\psi\rangle$ is defined as
\begin{equation}
\begin{aligned}
C(|\psi\rangle)=\sqrt{2(1-\textup{Tr}\rho_{1}^{2})},
\end{aligned}
\end{equation}
with $\rho_{1}=\textup{Tr}_{2}(|\psi\rangle\langle\psi|)$ being the reduced density matrix. For a mixed state $\rho$, the concurrence is given by
\begin{equation}
\begin{aligned}
C(\rho)=\min\limits_{\{p_{i},|\psi_{i}\rangle\}}\sum_{i}p_{i}C(|\psi_{i}\rangle),
\end{aligned}
\end{equation}
where the minimum is taken over all possible pure state decompositions $\rho=\sum_{i}p_{i}|\psi_{i}\rangle\langle \psi_{i}|$ with $p_{i}\geq 0$ and $\sum_{i}p_{i}=1$.

Let $F^{P}$ be the set of all separable pure states in $\mathcal{H}$. We use $\Gamma''$ to denote the set of the all local operations and classical communications. Obviously $\Gamma''$ is the subset of $\Gamma$. $\Theta''$ is defined as the set of the superchannels $\Theta''=\{\Phi_{\Gamma}''(N)=WNV|N\in L(\mathcal{H}),W,V\in\Gamma''\}$. It is easy to see that $\Theta''$ is the subset of $\Theta$.

To measure entanglement of channels, in addition to directly considering the difference between channel and all free channels, we can also study the transformation of all the separable states under a certain channel \cite{35}.

Definition $2$. For any quantum channel $N\in L(\mathcal{H})$, we define the entanglement measure of channels based on the concurrence as
\begin{equation}
\begin{aligned}
\Re_{c}(N)=\max\limits_{\rho\in F}C(N(\rho)).
\end{aligned}
\end{equation}
It is straightforward to get that the result is the same when the maximum is taken over all separable pure states, due to the convexity of concurrence. That is to say, this measure also can be expressed as
\begin{equation}
\begin{aligned}
\Re_{c}(N)=\max\limits_{|\psi\rangle\in F^{P}}C(N(|\psi\rangle)).
\end{aligned}
\end{equation}

It has the obvious advantage of being easy to calculate. Based on the properties of concurrence, for the entanglement measure $\Re_{c}$ one can obtain the following results.

$(X1)$ $(Nonnegativity)$ $\Re_{c}(N)\geq 0$, for any $N\in L(\mathcal{H})$; and $\Re_{c}(N)=0$ if and only if $N\in\Gamma$.

$(X2)$ $(Monotonicity)$ $\Re_{c}(\Phi_{\Gamma}''(N))\leq\Re_{c}(N)$, for any $\Phi_{\Gamma}''\in\Theta''$.

$Proof.$ We find
\begin{align*}
\Re_{c}(\Phi_{\Gamma}''(N))=\max\limits_{\rho\in F}C(WNV(\rho))\leq\max\limits_{\rho\in F}C(WN(\rho))\leq\max\limits_{\rho\in F}C(N(\rho))=\Re_{c}(N),
\end{align*}
where the first inequality follows from the set $F$ containing the set $V(F)=\{V(\rho)|\rho\in F\}$, the second inequality is due to the monotonicity of the concurrence.

$(X3)$ $(Convexity)$ $\Re_{c}(\sum\limits_{i}p_{i}N_{i})\leq\sum\limits_{i}p_{i}\Re_{c}(N_{i})$, where $N_{i}$ is quantum channel belonging to $L(\mathcal{H})$, and $\{p_{i}\}$ is a probability distribution satisfying $p_{i}\geq 0$ and $\sum_{i}p_{i}=1$.

$Proof.$ Obviously, we have
\begin{align*}
\Re_{c}(\sum\limits_{i}p_{i}N_{i})&=\max\limits_{\rho\in F}C(\sum\limits_{i}p_{i}N_{i}(\rho))\\
                                  &\leq\max\limits_{\rho\in F}\sum\limits_{i}p_{i}C(N_{i}(\rho))\\
                                  &\leq\sum\limits_{i}p_{i}\max\limits_{\rho\in F}C(N_{i}(\rho))\\
                                  &=\sum\limits_{i}p_{i}\Re_{c}(N_{i}),
\end{align*}
where the first inequality is obtained from the convexity of the concurrence, the second inequality comes from the nonnegativity of the concurrence.

$(X4)$ $(Strong$ $monotonicity)$ $\sum\limits_{i}p_{i}\Re_{c}(N_{i})\leq\Re_{c}(N)$, where $N_{i}=\Phi_{\Gamma}^{i''}(N)$, $\Phi_{\Gamma}^{i''}\in\Theta''$, and $\{p_{i}\}$ is a probability distribution satisfying $p_{i}\geq 0$ and $\sum_{i}p_{i}=1$.

$Proof.$ Evidently we have the following result
\begin{align*}
\sum\limits_{i}p_{i}\Re_{c}(N_{i})&=\sum\limits_{i}p_{i}\max\limits_{\rho\in F}C(N_{i}(\rho))\\
                                  &=\sum\limits_{i}p_{i}\max\limits_{\rho\in F}C(\Phi_{\Gamma}^{i''}(N)(\rho))\\
                                  &\leq\sum\limits_{i}p_{i}\max\limits_{\rho\in F}C(N(\rho))\\
                                  &=\Re_{c}(N),
\end{align*}
where the inequality holds because of the monotonicity $(X2)$ of entanglement measure $\Re_{c}$.

The origin of property $(X1)$ is that the free channels are known to contain no entanglement resource, i.e. they cannot increase entanglement of states. The property $(X2)$ is necessary, that is to say that the entanglement generating and increasing power of channels cannot be increased by the free superchannel in $\Theta''$. The properties $(X3),(X4)$ mean that the measure $\Re_{c}$ is convex and strong monotone. Next we will provide two examples to illustrate the entanglement measure $\Re_{c}$.

Example $1$. In the two bits system $\mathcal{H}=\mathcal{H}_{1}\otimes \mathcal{H}_{2}$ with dim$(\mathcal{H}_{i})=2$, $i=1,2$. Let consider a 2-qubit CNOT gate \cite{115}
\begin{equation}
\begin{aligned}
U_{\textup{CNOT}}=\sum\limits_{i=0}^1\sum\limits_{j=0}^1|i\rangle\langle i|\otimes|j\rangle\langle (i+j)\textup{mod}~2|.
\end{aligned}
\end{equation}

Let $\rho'=\frac{1}{2}(|0\rangle\langle0|+|0\rangle\langle1|+|1\rangle\langle0|+|1\rangle\langle1|)
\otimes|0\rangle\langle0|$. One can get $U_\textup{CNOT}(\rho')=\frac{1}{2}(|00\rangle\langle00|+|00\rangle\langle11|+|11\rangle\langle00|
+|11\rangle\langle11|)$ which is a maximally entangled state.

Then, $\Re_{c}(U_\textup{CNOT})=\max\limits_{\rho\in F}C(U_\textup{CNOT}(\rho))=C(U_\textup{CNOT}
(\rho'))=1$.

For any quantum state $\sigma\in D(\mathcal{H})$, the value $C(\sigma)\in[0,1]$. According to the definition of entanglement measure $\Re_{c}$, one can obtain $\Re_{c}(N)\in[0,1]$, for any quantum channel $N\in L(\mathcal{H})$. Therefore, $U_\textup{CNOT}$ is a meaningful entanglement resource in the quantum channel. In the study of the relations between quantum coherence and quantum entanglement and quantum nonlocality \cite{115,10}, the entanglement of incoherent channels is also worth to consider in addition to coherent resource of quantum states. The channel $U_{\textup{CNOT}}$ can convert the maximum coherent state to the maximum entangled state, which is impossible for any channel with entanglement measure $\Re_{c}$ less than 1.

Example $2$. Consider the 2-qubit quantum channel
\begin{equation}
\begin{aligned}
N=\sum\limits_{i=0}^1|i\rangle\langle (i+1)\textup{mod}~2|\otimes|(i+1)\textup{mod}~2\rangle\langle i|+\sum\limits_{j=0}^1|j\rangle\langle j|\otimes|j\rangle\langle j|.
\end{aligned}
\end{equation}

In the system $\mathcal{H}=\mathcal{H}_{1}\otimes \mathcal{H}_{2}$ with dim$(\mathcal{H}_{i})=2$, $i=1,2$, for any separable pure state
\begin{equation}
\begin{aligned}
\rho=\begin{pmatrix} 1-a & \sqrt{a(1-a)}\textup{e}^{-\textup{i}\phi} \\ \sqrt{a(1-a)}\textup{e}^{\textup{i}\phi} & a \end{pmatrix}\otimes\begin{pmatrix} 1-a' & \sqrt{a'(1-a')}\textup{e}^{-\textup{i}\phi'} \\ \sqrt{a'(1-a')}\textup{e}^{\textup{i}\phi'} & a' \end{pmatrix},
\end{aligned}
\end{equation}
where $a,a'\in[0,1]$, and $\phi,\phi'\in[0,2\pi]$, we get
\begin{equation}
\begin{aligned}
N(\rho)=\begin{pmatrix} 1-a' & \sqrt{a'(1-a')}\textup{e}^{-\textup{i}\phi'} \\ \sqrt{a'(1-a')}\textup{e}^{\textup{i}\phi'} & a' \end{pmatrix}\otimes\begin{pmatrix} 1-a & \sqrt{a(1-a)}\textup{e}^{-\textup{i}\phi} \\ \sqrt{a(1-a)}\textup{e}^{\textup{i}\phi} & a \end{pmatrix}.
\end{aligned}
\end{equation}
It is easy to know $\Re_{c}(N)=\max\limits_{\rho\in F^{P}}C(N(\rho))=\max\limits_{\rho\in F^{P}}C(\rho)=0$.

The action of this quantum channel is to swap the particles of the two subsystems. It obviously cannot increase entanglement of separable quantum states.

\subsection{The measure based on the $k$-ME concurrence}
In the multipartite Hilbert space $\mathcal{H}=\mathcal{H}_{1}\otimes \mathcal{H}_{2}\otimes\cdots\otimes\mathcal{H}_{n}$ with dim$(\mathcal{H}_{i})=d_{i}$, $i=1,2,\ldots,n$, for the pure state $|\psi\rangle$, the $k$-ME concurrence \cite{86,15,12,112} is defined as
\begin{equation}
\begin{aligned}
C_{k-\mathrm{ME}}(|\psi\rangle)=\min\limits_{A}\sqrt{2\left(1-\frac{\sum\limits_{t=1}^k\textrm{Tr}(\rho^2_{A_t})}{k}\right)}
=\min\limits_{A}\sqrt{\frac{2\sum\limits_{t=1}^k\left[1-\textrm{Tr}(\rho^2_{A_t})\right]}{k}},
\end{aligned}
\end{equation}
where $\rho=|\psi\rangle\langle\psi|$, the reduced density matrix $\rho_{A_t}$ is obtained by tracing over the subsystem $\bar{A_t}$ ($\bar{A_t}$ is the complement of $A_t$ in $\{1,2,\ldots,n\}$) and the minimum is taken over all possible $k$-partitions $A=A_1|A_2|\cdots|A_k$ of $\{1,2,\ldots,n\}$.

For the mixed state $\rho$, the $k$-ME concurrence is given as the convex roof,
\begin{equation}
\begin{aligned}
C_{k-\mathrm{ME}}(\rho)=\inf\limits_{\{p_{m},|\psi_{m}\rangle\}}\sum\limits_{m}p_{m}C_{k-\mathrm{ME}}(|\psi_{m}\rangle),
\end{aligned}
\end{equation}
of all possible pure state decompositions $\rho=\sum_{m}p_{m}|\psi_{m}\rangle\langle\psi_{m}|$.

Concurrence is a special case of $k$-ME concurrence. The measure $\Re_{c}$ is a useful entanglement measure of channels with respect to two parties. In multipartite systems, we use the $k$-ME concurrence to define the entanglement measure of multipartite quantum channels.

Definition $3$. For any quantum channel $N\in L(\mathcal{H})$, the entanglement measure of multipartite quantum channels based on $k$-ME concurrence is defined as
\begin{equation}
\begin{aligned}
\Re_{k-\mathrm{ME}}(N)=\max\limits_{\rho\in F}C_{k-\mathrm{ME}}(N(\rho)).
\end{aligned}
\end{equation}

Because of the convexity of $k$-ME concurrence, we can narrow the range of separable states researched to all fully separable pure states. The entanglement measure based on $k$-ME concurrence can also be expressed as
\begin{equation}
\begin{aligned}
\Re_{k-\mathrm{ME}}(N)=\max\limits_{|\psi\rangle\in F^{P}}C_{k-\mathrm{ME}}(N(|\psi\rangle)).
\end{aligned}
\end{equation}

As the number of subsystems increases, so does the complexity involved in quantifying the entanglement of channels. In the $n$-partite Hilbert space $\mathcal{H}$, it is possible that a quantum channel cannot transform fully separable states to genuine entangled states but it can transform a fully separable state to a $k$-nonseparable $(2<k<n)$ state. In this paper, the concept of $k$-nonseparable is reinforced, a quantum state is called $S$-$k$-nonseparable $(2<k\leq n)$ if the state is nonseparable in every $k$-partition and separable in some $(k-1)$-partitions. In particular, if a state is a genuine entangled state, then it is called $S$-$2$-nonseparable. Let $S_{N}$ be the set $\{N(\rho)|\rho\in F^{P}\}$, where $N$ is any quantum channel in $L(\mathcal{H})$. Let $K(\rho')=k$ $(2\leq k\leq n)$, if the state $\rho'$ is $S$-$k$-nonseparable; otherwise, $K(\rho')=n+1$. The strength value of the channel $N$ is defined as
\begin{equation}
\begin{aligned}
K(N)=\min\limits_{\rho'\in S_{N}}K(\rho'),
\end{aligned}
\end{equation}
where the minimum is taken over all $\rho'$ belonging to $S_{N}$.

According to the strength value of channel, the entanglement increasing power of quantum channel can be roughly classified. $Q_{k}$ $(2\leq k\leq n+1)$ denotes a set of quantum channel whose strength value is equal to $k$. Specially, the element of the set $Q_{2}$ can convert some fully separable states into genuine entangled states and the channel in the set $Q_{n+1}$ cannot increase the entanglement of any fully separable quantum states. It is obvious that the channel in the set $Q_{n+1}$ is free channel. We can get that the powers of the channel sets to generate the entanglement  are in the following order
\begin{equation}
\begin{aligned}
Q_{2}\succ\cdots\succ Q_{k}\succ\cdots\succ Q_{n+1},~~k=3,\ldots,n.
\end{aligned}
\end{equation}

For every $k$ $(2< k\leq n+1)$, if $N\in Q_{k}$, then $\Re_{(k-i)-\mathrm{ME}}(N)=0$ for $i=1,\ldots,k-2$, and $\Re_{k-\mathrm{ME}}(N)>0$. The values of entanglement measure are strictly positive for the channels in $Q_{2}.$ Let $\Gamma_{k}$ $(2\leq k\leq n)$ be the set of channels which convert any fully separable state into a $k$-separable state. It is not difficult to find that the  powers of the channel sets to generate the entanglement  are in the  order
\begin{equation}
\begin{aligned}
\Gamma_{2}\succ\cdots\succ\Gamma_{k}\succ\cdots\succ\Gamma_{n},~~k=3,\ldots,n-1.
\end{aligned}
\end{equation}
When $k=n$, the set $\Gamma_{k}$ is actually $\Gamma$.

Based on the properties of $k$-ME concurrence, the entanglement measure $\Re_{k-\mathrm{ME}}$ satisfies the following properties.

$(V1)$ $(Nonnegativity)$ $\Re_{k-\mathrm{ME}}(N)\geq 0$, for any $N\in L(\mathcal{H})$; and $\Re_{k-\mathrm{ME}}(N)=0$ if and only if $N\in\Gamma_{k}$.

$(V2)$ $(Monotonicity)$ $\Re_{k-\mathrm{ME}}(\Phi_{\Gamma}''(N))\leq\Re_{k-\mathrm{ME}}(N)$, for any $\Phi_{\Gamma}''\in\Theta''$.

$(V3)$ $(Convexity)$ $\Re_{k-\mathrm{ME}}(\sum\limits_{i}p_{i}N_{i})\leq\sum\limits_{i}p_{i}\Re_{k-\mathrm{ME}}(N_{i})$, where $N_{i}$ is a quantum channel belonging to $L(\mathcal{H})$, and $\{p_{i}\}$ is a probability distribution satisfying $p_{i}\geq 0$ and $\sum_{i}p_{i}=1$.

$(V4)$ $(Strong$ $monotonicity)$ $\sum\limits_{i}p_{i}\Re_{k-\mathrm{ME}}(N_{i})\leq\Re_{k-\mathrm{ME}}(N)$, where $N_{i}=\Phi_{\Gamma}^{i''}(N)$, $\Phi_{\Gamma}^{i''}\in\Theta''$, and $\{p_{i}\}$ is a probability distribution satisfying $p_{i}\geq 0$ and $\sum_{i}p_{i}=1$.

$(V5)$ $(Subadditivity)$ $\Re_{k-\mathrm{ME}}(N\otimes M)\leq \Re_{k-\mathrm{ME}}(N)+\Re_{k-\mathrm{ME}}(M)$, for any $N,M \in L(\mathcal{H})$.

$Proof.$ Obviously, we have
\begin{align*}
\Re_{k-\mathrm{ME}}(N\otimes M)&=\max\limits_{\rho\in F^{\otimes 2}}C_{k-\mathrm{ME}}(N\otimes M(\rho))\\
                               &=\max\limits_{\rho_{1}\otimes \rho_{2}\in F^{\otimes 2}}C_{k-\mathrm{ME}}(N\otimes M(\rho_{1}\otimes \rho_{2}))\\
                               &\leq \max\limits_{\rho_{1}\otimes \rho_{2}\in F^{\otimes 2}}[C_{k-\mathrm{ME}}(N(\rho_{1}))+C_{k-\mathrm{ME}}(M(\rho_{2}))]\\
                               &=\max\limits_{\rho_{1}\in F}C_{k-\mathrm{ME}}(N(\rho_{1}))+\max\limits_{\rho_{2}\in F}C_{k-\mathrm{ME}}(M(\rho_{2}))\\
                               &=\Re_{k-\mathrm{ME}}(N)+\Re_{k-\mathrm{ME}}(M),
\end{align*}
where the inequality is derived from the subadditivity of $k$-ME concurrence \cite{86}, the third equality comes from the nonnegativity of $k$-ME concurrence and the mutual independence of choosing $\rho_{i}$ for $i=1,2$.

The measure $\Re_{k-\mathrm{ME}}$ can quantify the entanglement of channels in multipartite system. It meets the natural requirements of the measure. Two specific examples are given to explain the measure $\Re_{k-\mathrm{ME}}$.

Example $3$. Consider the $n$-qudit unitary operation
\begin{equation}
\begin{aligned}
U=\sum\limits_{i_{1}=0}^{d-1}\sum\limits_{i_{2}=0}^{d-1}\cdots \sum\limits_{i_{n}=0}^{d-1}|i_{1}\rangle\langle i_{1}|\otimes|(i_{1}+i_{2})\textup{mod}~d\rangle\langle i_{2}|\otimes\cdots\otimes|(i_{1}+i_{n})\textup{mod}~d\rangle\langle i_{n}|.
\end{aligned}
\end{equation}

An $n$-qudit fully separable quantum state $\rho'=\rho^{1}\otimes|0\rangle\langle0|^{2}\otimes\cdots\otimes|0\rangle\langle0|^{n}$ with $\rho^{1}=\frac{1}{d}\sum_{ij}|i\rangle\langle j|$, will be transform to the quantum state $U(\rho')=\frac{1}{d}\sum_{ij}|ii\cdots i\rangle_{n}\langle jj\cdots j|$ by $U$. It is a genuinely maximally entangled state. The strength value of the channel $U$ is 2.

Then, it is easy to obtain $\Re_{2-\mathrm{ME}}(U)=\max\limits_{\rho\in F}C_{2-\mathrm{ME}}(U(\rho))=C_{2-\mathrm{ME}}(U(\rho'))=\sqrt{\frac{2(d-1)}{d}}$.

The relationship between coherent and tripartite entangled resources is established in \cite{10}. It can also be considered to construct the relationship between coherent and multipartite entangled resources under the incoherent channel with entanglement. To this, the influence of the incoherent channel with different degree of entanglement is also worth studying. As can be seen, the channel $U$ has important meaning. $\rho^{1}$ is a maximally coherent state. Under the channel $U$, the state $\rho'$ is converted to a maximally multipartite entangled state.

Example $4$. Consider an $n$-qudit quantum channel
\begin{equation}
\begin{aligned}
M=\sum\limits_{i_{1}i_{2}\cdots i_{n}}^{d-1}|i_{n}\rangle\langle i_{1}|\otimes|i_{1}\rangle\langle i_{2}|\otimes\cdots \otimes|i_{n-1}\rangle\langle i_{n}|.
\end{aligned}
\end{equation}

For any fully separable pure state $\rho=\rho_{1}\otimes\rho_{2}\otimes\cdots \otimes\rho_{n}$, $\rho_{i}$ is any pure state in $i$-th subsystem, we have
$M(\rho)=\rho_{n}\otimes\rho_{1}\otimes\cdots \otimes\rho_{n-1}$. This result shows that the channel only changes the position of the particles, cannot increase the overall entanglement. Thus, the strength value of channel $M$ is $n+1$. It belongs to $Q_{n+1}$. Naturally, $\Re_{k-\mathrm{ME}}(M)=0$ for $k=2,\ldots,n$, the entanglement of channel $M$ is vanish. Obviously, it has no value in the question of converting coherent states to entangled states.

\section{Conclusion}
In this paper, we introduce some concepts about free quantum channels and free quantum superchannels, and have investigated the generalized entanglement of channels. The main advance is to provide some well-defined entanglement measures ($\Re_{r}$, $\Re_{c}$, $\Re_{k-\mathrm{ME}}$) based on the Choi relative entropy of quantum channels, concurrence and $k$-ME concurrence respectively. On the one hand, the entanglement measure $\Re_{r}$ directly quantifies the difference between the channel and all free channels. On the other hand, the entanglement measure $\Re_{c}$ and $\Re_{k-\mathrm{ME}}$ quantify the entanglement of free state conversion under the quantum channel. They are convex and strong monotone. Furthermore, $\Re_{r}$ is additive and $\Re_{k-\mathrm{ME}}$ is subadditive. In the multipartite system, the measure $\Re_{k-\mathrm{ME}}$ is considered more carefully and comprehensively. In order to characterize the entanglement measures more clearly, several examples of calculating the entanglement of quantum channel are given. These examples also show that the research of channels has certain value in the study of the conversions between quantum coherence states and multipartite entanglement, nonlocality states. We hope these results will give more insight into the channel.

\begin{acknowledgments}
This project was supported by the National Natural Science Foundation of China under Grant No. 12071110, the Hebei Natural Science Foundation of China under Grant No. A2020205014, and the Education Department of Hebei Province Natural Science Foundation Grant Nos. ZD2020167, ZD2021066.
\end{acknowledgments}


\begin{thebibliography}{99}




%1-10
\bibitem{Horodecki} R. Horodecki, P. Horodecki, M. Horodecki, and K. Horodecki, Quantum entanglement, \href{https://journals.aps.org/rmp/abstract/10.1103/RevModPhys.81.865} {Rev. Mod. Phys. \textbf{81}, 865 (2009)}.
\bibitem{40} B. Groisman and S. Strelchuk, Optimal amount of entanglement to distinguish quantum states instantaneously, \href{https://journals.aps.org/pra/abstract/10.1103/PhysRevA.92.052337} {Phys. Rev. A \textbf{92}, 052337 (2015)}.
\bibitem{43} S. Rout, A. G. Maity, and A. Mukherjee, Genuinely nonlocal product bases: Classification and entanglement-assisted discrimination, \href{https://journals.aps.org/pra/abstract/10.1103/PhysRevA.100.032321} {Phys. Rev. A \textbf{100}, 032321 (2019)}.
\bibitem{20} S. M. Cohen, Understanding entanglement as resource: Locally distinguishing unextendible product bases, \href{https://journals.aps.org/pra/abstract/10.1103/PhysRevA.77.012304} {Phys. Rev. A \textbf{77}, 012304 (2008)}.
\bibitem{PRL70.1895} C. H. Bennett, G. Brassard, C. Cr\'{e}peau, R. Jozsa, A. Peres, and W. K. Wootters, Teleporting an unknown quantum state via dual classical and Einstein-Podolsky-Rosen channels, \href{https://journals.aps.org/prl/abstract/10.1103/PhysRevLett.70.1895} {Phys. Rev. Lett. \textbf{70}, 1895 (1993)}.
\bibitem{PR448.1} X. B. Wang, et al, Quantum information with Gaussian states, \href{https://doi.org/10.1016/j.physrep.2007.04.005} {Phys. Rep. \textbf{448}, 1 (2007)}.
\bibitem{18} Y. H. Luo, H. S. Zhong, and X. L. Wang, Quantum teleportation in high dimensions, \href{https://journals.aps.org/prl/abstract/10.1103/PhysRevLett.123.070505} {Phys. Rev. Lett. \textbf{123}, 070505 (2019)}.
\bibitem{Science283.2050} H. K. Lo and H. F. Chau, Unconditional security of quantum key distribution over arbitrarily long distances, \href{https://science.sciencemag.org/content/283/5410/2050} {Science \textbf{283}, 2050 (1999)}.
\bibitem{PRL67.661} A. K. Ekert, Quantum cryptography based on Bell's theorem, \href{https://journals.aps.org/prl/abstract/10.1103/PhysRevLett.67.661} {Phys. Rev. Lett. \textbf{67}, 661 (1991)}.
\bibitem{PRL68.557} C. H. Bennett, G. Brassard, and N. D. Mermin, Quantum cryptography without Bell's theorem, \href{https://journals.aps.org/prl/abstract/10.1103/PhysRevLett.68.557} {Phys. Rev. Lett. \textbf{68}, 557 (1992)}.

%11-20
\bibitem{57} V. Vedral and M. B. Plenio, Entanglement measures and purification procedures, \href{https://journals.aps.org/pra/abstract/10.1103/PhysRevA.57.1619} {Phys. Rev. A \textbf{57}, 1619 (1998)}.
\bibitem{78} V. Vedral, M. B. Plenio, M. A. Rippin, and P. L. Knight, Quantifying entanglement, \href{https://journals.aps.org/prl/abstract/10.1103/PhysRevLett.78.2275} {Phys. Rev. Lett. \textbf{78}, 2275 (1997)}.
\bibitem{80} W. K. Wootters, Entanglement of formation of an arbitrary state of two qubits, \href{journals.aps.org/prl/abstract/10.1103/PhysRevLett.80.2245} {Phys. Rev. Lett. \textbf{80}, 2245 (1998)}.
\bibitem{86} Y. Hong, T. Gao, and F. L. Yan, Measure of multipartite entanglement with computable lower bounds, \href{journals.aps.org/pra/abstract/10.1103/PhysRevA.86.062323} {Phys. Rev. A \textbf{86}, 062323 (2012)}.
\bibitem{58} K. Zyczkowski, P. Horodecki, A. Sanpera, and M. Lewenstein, Volume of the set of separable states, \href{https://journals.aps.org/pra/abstract/10.1103/PhysRevA.58.883} {Phys. Rev. A \textbf{58}, 883 (1998)}.
\bibitem{65} G. Vidal and R. F. Werner, Computable measure of entanglement, \href{journals.aps.org/pra/abstract/10.1103/PhysRevA.65.032314} {Phys. Rev. A \textbf{65}, 032314 (2002)}.
\bibitem{54} C. H. Bennett, D. P. DiVincenzo, J. A. Smolin, and W. K. Wootters, Mixed-state entanglement and quantum error correction, \href{https://journals.aps.org/pra/abstract/10.1103/PhysRevA.54.3824} {Phys. Rev. A \textbf{54}, 3824 (1996)}.
\bibitem{246} P. W. Shor, Equivalence of additivity questions in quantum information theory, \href{https://link.springer.com/article/10.1007/s00220-003-0981-7} {Commun. Math. Phys. \textbf{246}, 453 (2004)}.
\bibitem{Baumgratz} T. Baumgratz, M. Cramer, and M. B. Plenio, Quantifying coherence, \href{https://journals.aps.org/prl/abstract/10.1103/PhysRevLett.113.140401} {Phys. Rev. Lett. \textbf{113}, 140401 (2014)}.
\bibitem{Chitambar} E. Chitambar and G. Gour, Quantum resource theories, \href{https://journals.aps.org/rmp/abstract/10.1103/RevModPhys.91.025001} {Rev. Mod. Phys. \textbf{91}, 025001 (2019)}.


%21-30
\bibitem{122} C. T. Patricia, et al, Resource theory of entanglement with a unique multipartite maximally entangled state, \href{journals.aps.org/prl/abstract/10.1103/PhysRevLett.122.120503} {Phys. Rev. Lett. \textbf{122}, 120503 (2019)}.
\bibitem{41} Y. C. Liu and X. Yuan, Operational resource theory of quantum channels, \href{https://journals.aps.org/prresearch/abstract/10.1103/PhysRevResearch.2.012035} {Phys. Rev. Research \textbf{2}, 012035(R) (2020)}.
\bibitem{37} T. Theurer, D. Egloff, L. Zhang, and M. B. Plenio, Quantifying operations with an application to coherence, \href{https://journals.aps.org/prl/abstract/10.1103/PhysRevLett.122.190405} {Phys. Rev. Lett. \textbf{122}, 190405 (2019)}.
\bibitem{38} Z. W. Liu and A. Winter, Resource theories of quantum channels and the universal role of resource erasure, \href{https://arxiv.org/abs/1904.04201} {arXiv:1904.04201}.
\bibitem{123} G. Gour and A. Winter, How to quantify a dynamical quantum resource, \href{https://journals.aps.org/prl/abstract/10.1103/PhysRevLett.123.150401} {Phys. Rev. Lett. \textbf{123}, 150401 (2019)}.
\bibitem{52} G. Gour and C. M. Scandolo, The entanglement of a bipartite channel, \href{https://arxiv.org/abs/1907.02552} {arXiv:1907.02552}.
\bibitem{181} S. B$\ddot{\textup{a}}$uml, S. Das, X. Wang, and M. M. Wilde, Resource theory of entanglement for bipartite quantum channels, \href{https://arxiv.org/abs/1907.04181} {arXiv:1907.04181}.

\bibitem{125} G. Gour and C. M. Scandolo, Dynamical entanglement, \href{https://journals.aps.org/prl/abstract/10.1103/PhysRevLett.125.180505} {Phys. Rev. Lett. \textbf{125}, 180505 (2020)}.
\bibitem{100} J. W. Xu, Coherence of quantum channels, \href{https://journals.aps.org/pra/abstract/10.1103/PhysRevA.100.052311} {Phys. Rev. A \textbf{100}, 052311 (2019)}.
\bibitem{35} L. Li, K. F. Bu, and Z. W. Liu, Quantifying the resource content of quantum channels: An operational approach, \href{https://journals.aps.org/pra/abstract/10.1103/PhysRevA.101.022335} {Phys. Rev. A \textbf{101}, 022335 (2020)}.
\bibitem{97} F. Leditzky, E. Kaur, N. Datta, and M. M. Wilde, Approaches for approximate additivity of the Holevo information of quantum channels, \href{https://journals.aps.org/pra/abstract/10.1103/PhysRevA.97.012332} {Phys. Rev. A \textbf{97}, 012332 (2018)}.
\bibitem{17} X. Yuan, Hypothesis testing and entropies of quantum channels, \href{https://journals.aps.org/pra/abstract/10.1103/PhysRevA.99.032317} {Phys. Rev. A \textbf{99}, 032317 (2019)}.
\bibitem{87} K. Audenaert, J. Eisert, E. Jan\'{e}, M. B. Plenio, S. Virmani, and B. De Moor, Asymptotic relative entropy of entanglement, \href{https://journals.aps.org/prl/abstract/10.1103/PhysRevLett.87.217902} {Phys. Rev. Lett. \textbf{87}, 217902 (2001)}.
\bibitem{42} S. Hill and W. K. Wootters, Entanglement of a pair of quantum bits, \href{https://journals.aps.org/prl/abstract/10.1103/PhysRevLett.78.5022} {Phys. Rev. Lett. \textbf{78}, 5022 (1997)}.


%31-37
\bibitem{21} P. Rungta, V. Buzek, C. M. Caves, M. Hillery, and G. J. Milburn, Universal state inversion and concurrence in arbitrary dimensions, \href{https://journals.aps.org/pra/abstract/10.1103/PhysRevA.64.042315} {Phys. Rev. A \textbf{64}, 042315 (2001)}.
\bibitem{22} F. Mintert, M. Kus, and A. Buchleitner, Concurrence of mixed bipartite quantum states in arbitrary dimensions, \href{https://journals.aps.org/prl/abstract/10.1103/PhysRevLett.92.167902} {Phys. Rev. Lett. \textbf{92}, 167902 (2004)}.
\bibitem{115} A. Streltsov, U. Singh, and H. S. Dhar, Measuring quantum coherence with entanglement, \href{https://journals.aps.org/prl/abstract/10.1103/PhysRevLett.115.020403} {Phys. Rev. Lett. \textbf{115}, 020403 (2015)}.
\bibitem{10} Y. Xi, T. G. Zhang, Z. J. Zheng, X. Q. Li-Jost, and S. M. Fei, Converting quantum coherence to genuine multipartite entanglement and nonlocality, \href{https://journals.aps.org/pra/abstract/10.1103/PhysRevA.100.022310} {Phys. Rev. A \textbf{100}, 022310 (2019)}.
\bibitem{12} T. Gao, Y. Hong, Y. Lu, and F. L. Yan, Efficient $k$-separability criteria for mixed multipartite quantum states, \href{https://iopscience.iop.org/article/10.1209/0295-5075/104/20007} {Europhys. Lett. \textbf{104}, 20007 (2013)}.
\bibitem{15} T. Gao and Y. Hong, Detection of genuinely entangled and nonseparable $n$-partite quantum states, \href{https://journals.aps.org/pra/abstract/10.1103/PhysRevA.82.062113} {Phys. Rev. A \textbf{82}, 062113 (2010)}.
\bibitem{112} T. Gao, F. L. Yan, and S. J. van Enk, Permutationally invariant part of a density matrix and nonseparability of $N$-qubit states, \href{https://journals.aps.org/prl/abstract/10.1103/PhysRevLett.112.180501} {Phys. Rev. Lett. \textbf{112}, 180501 (2014)}.












\end{thebibliography}
\end{document}